\documentclass[%
 reprint,
 amsmath,amssymb,
 aps,
citeautoscript]{revtex4-2}

\setcitestyle{super}

\usepackage{amssymb}
\usepackage{amsthm}
\usepackage{upgreek} 
\usepackage{graphicx}
\usepackage{dcolumn}
\usepackage{bm}
\usepackage[colorlinks=true,bookmarks=false,citecolor=black,urlcolor=blue,linkcolor=black]{hyperref} 

\usepackage{color}
\usepackage{soul}
\soulregister\cite{7} 
\soulregister\ref{7} 
\soulregister\eqref{7} 

\begin{document}

\title{Bi-stability and period-doubling cascade of frequency combs \\in exceptional-point lasers}

\author{Xingwei Gao$^1$}\email{xingweig@usc.edu}
\author{Hao He$^1$}
\author{Weng W. Chow$^2$}
\author{Alexander Cerjan$^2$}\email{awcerja@sandia.gov}
\author{Chia Wei Hsu$^1$}

\affiliation{
$^1$Ming Hsieh Department of Electrical and Computer Engineering, University of Southern California, Los Angeles, CA 90007, USA \\
$^2$Center for Integrated Nanotechnologies, Sandia National Laboratories, Albuquerque, NM 87185, USA
}

\begin{abstract}
Recent studies have demonstrated that a laser can self-generate frequency combs when tuned near an exceptional point (EP), where two cavity modes coalesce. These EP combs induce periodic modulation of the population inversion in the gain medium, and their repetition rate is independent of the laser cavity's free spectral range. In this work, we perform a stability analysis that reveals two notable properties of EP combs, bi-stability and a period-doubling cascade. The period-doubling cascade enables halving of the repetition rate while maintaining the comb's total bandwidth, presenting opportunities for the design of highly compact frequency comb generators.
\end{abstract}

\maketitle

\section{Introduction}\label{Intro}
\vspace{-4pt}
A frequency comb is an optical phenomenon where a system produces a series of equally-spaced spectral lines. Optical frequency combs (OFCs) are essential to optical-frequency synthesizers~\cite{spencer2018optical} and precision metrology~\cite{collaboration2021frequency, Leopardi:17, doi:10.1126/science.1154622}. In the past decade, OFCs have also been applied in optical communications~\cite{marin2017microresonator} and quantum computation~\cite{roslund2014wavelength}. Conventionally, OFCs are generated by mode locked lasers~\cite{RevModPhys.75.325}, optical cavities with nonlinearity ~\cite{2007_DelHaye_Nature,2018_Kippenberg_Science_review,Parriaux:20}, and quantum cascade lasers~\cite{hugi2012mid, silvestri2023frequency,2024_Opacak_Nature}. However, these traditional comb generation methods all require the repetition rate to match the free spectral range (FSR) of the laser cavity. Consequently, large cavity sizes are required to generate radio-frequency OFCs. 

Recently, it was discovered that exceptional point (EP) optical cavities can develop into frequency combs with repetition rates independent of the cavity's FSR~\cite{PhysRevA.107.033509,Gao2024PALT}.
EPs are points of degeneracy in the phase space where two or more eigenmodes become identical~\cite{2011_Moiseyev_book,Heiss_2012,feng2017non, el2018non, doi:10.1126/science.aar7709}. When two cavity modes are sufficiently close to an EP, there exists a gain threshold above which any perturbation to the system induces a long-lived periodic modulation to the carrier populations in the gain medium. These dynamic populations subsequently modulate any active lasing modes in the system, generating equally spaced comb lines in the output spectrum. As such, an EP comb is self-generated without any external modulation, and its repetition rate $\omega_\text{d}$, equal to the self-modulation rate of the inversion, can be significantly smaller than the cavity's free spectral range as it is approximately set by the frequency spacing of the EP modes.
In principal, arbitrarily small repetition rates can be achieved as the laser system is tuned sufficiently close to the EP. However, in practice, it is technically difficult to reach an exact EP in experiments. Moreover, the robustness of the comb will be compromised due to the enhanced sensitivity near EP~\cite{doi:10.1126/science.aar7709,2022_Benzaouia_APLph,Gao2024PALT}. Therefore, the minimum achievable repetition rate in EP combs is thought to be limited.  

In this work, we show that one can significantly reduce the repetition rate of EP combs without pushing the system closer to the EP, thus realizing robust radio-frequency OFCs in small laser cavities. Specifically, we demonstrate that 
an EP comb can halve its repetition rate multiple times through a period-doubling cascade, which was typically observed in more complicated laser systems with light injection or an external modulation.\cite{PhysRevA.45.1893,PhysRevA.51.4181}
In particular, we carry out a stability analysis with the perturbation method on the Maxwell-Bloch equations~\cite{haken1985laser, PhysRevA.54.3347}. Driven by the periodic population inversion, an EP laser becomes a Floquet system \cite{annurev-conmatphys-040721-015537}, where any infinitesimal perturbation can be decomposed into Floquet eigenmodes, with each having a complex Floquet frequency. 
A Floquet frequency with a positive (negative) imaginary part leads to the corresponding Floquet mode growing (decaying) over time.   
The stability of an EP comb is thus determined by 
the sign of $\text{Im}(\omega_\text{F})$, with $\omega_\text{F}$ being the Floquet frequency with the largest imaginary part.
Given an existing EP comb with a line spacing of $\omega_\text{d}$,
by solving for $\omega_{\textrm{F}}$, we find a series of pumping thresholds at which 
a Floquet mode turns on, with $\text{Im}(\omega_{\text{F}})=0$ and $\text{Re}(\omega_{\text{F}})=0.5\omega_\text{d}$. Through gain saturation, this Floquet mode induces an additional modulation to the population inversion, which has twice the period of the inversion fluctuation from the original EP comb. The re-modulated inversion then doubles the period of the lasing field's envelope, hence inserting extra lines into the original EP comb.
The period doubling occurs through each of these thresholds cascade, eventually leading to arbitrarily small repetition rate.
With the perturbation method, we also find a bistability zone in EP lasers, where two different EP combs exist at the same pumping strength. Thus, the laser state depends on initialization,
which is potentially applicable in optical signal-processing devices and all-optical computer systems.\cite{siegman1986lasers}


\section{Stability Analysis on Frequency Combs}\label{sec:SA}
\vspace{-4pt}

Period doubling is a special transition from one frequency comb to another. For the transition to occur, the former comb must first become unstable. To predict the stability of a comb, we apply the first-order perturbation method to the fundamental equations governing lasing materials. We begin by deriving generalized perturbation equations for arbitrary lasing states. 
Then, we extend the analysis to limit-cycle lasing states, proving that the stability of a frequency comb is associated with a complex Floquet frequency $\omega_{\text{F}}$, which can be determined by solving a linear eigenmode equation.

Lasers can be described rigorously by the Maxwell-Bloch (MB) equations~\cite{haken1985laser, PhysRevA.54.3347, Gao2024PALT}, a semi-classical model depicting the relations among the population inversion $D(\mathbf{r},t)$ of gain media, the electric field $\mathbf{E}(\mathbf{r},t)$ and the polarization density $\mathbf{P}(\mathbf{r},t)$. To simplify the notation, we focus on a one-dimensional (1D) laser cavity with $\mathbf{E}(\mathbf{r},t)=E(x,t)\hat{\mathbf{z}}$, $\mathbf{P}(\mathbf{r},t)=P(x,t)\hat{\mathbf{z}}$ and $D(\mathbf{r},t)=D(x,t)$ (Our method can be readily extended to three-dimensional systems, as shown in Supplementary section I). In particular, the MB equations are
\begin{align}
&\frac{\partial}{\partial t}D = - \gamma_{\parallel}(D-D_\text{p})-\frac{i\gamma_{\parallel}}{2}(E^*P-EP^*), \label{MB:D}\\
&\frac{\partial}{\partial t}{P} = -(i\omega_{ba} + \gamma_{\perp}){P}-i\gamma_{\perp}DE, \label{MB:P} \\
&\frac{\partial^2 E}{\partial x^2} - \frac{1}{c^2} \left(\varepsilon_c\frac{\partial^2}{\partial t^2}+\frac{\sigma}{\varepsilon_0} \frac{\partial}{\partial t}\right)E = \frac{1}{c^2} \frac{\partial^2}{\partial t^2} P \label{MB:E}.
\end{align}
$D$, $E$, and $P$ here have been normalized by $R^2/(\varepsilon_0\hbar\gamma_{\perp})$, $2R/(\hbar\sqrt{\gamma_{\perp}\gamma_\parallel})$, and $2R/(\varepsilon_0\hbar\sqrt{\gamma_{\perp}\gamma_\parallel})$, respectively,
with $R$ being the amplitude of the atomic dipole moment, $\varepsilon_0$ the vacuum permittivity, $\hbar$ the Planck constant, and $\gamma_\perp$ the dephasing rate of the gain-induced polarization ({\it i.e.}, the bandwidth of the gain).
$D_\text{p}(r)$ is the normalized net pumping strength and profile, $\omega_{ba}$ is the frequency gap between the two atomic levels, $\varepsilon_c(x)$ is the relative permittivity profile of the cold cavity, $\sigma(x)$ is a conductivity profile that produces linear absorption, and $c$ is the vacuum speed of light. 

Consider a fixed-point or limit-cycle solution to the MB equations (\ref{MB:D}--\ref{MB:E}),
$D=D_\text{s}(x,t)$, $E=E_\text{s}(x,t)$ and $P=P_\text{s}(x,t)$. To determine the stability of the solution, we add a small perturbation, such that $D=D(x,t)_\text{s}+\Delta d(x,t)$, $E=E_\text{s}(x,t)+\Delta \epsilon(x,t)$, and $P=P_\text{s}(x,t)+\Delta p(x,t)$, where $\Delta$ is a real infinitesimal number. The perturbation equations are derived by substituting the perturbed $D$, $P$ and $E$ into Eqs.~(\ref{MB:D})--(\ref{MB:E}) and then extracting the linear terms of $\Delta$,
\begin{align}
&\frac{\partial}{\partial t}d = - \gamma_{\parallel}d-\frac{i\gamma_{\parallel}}{2}(\{E_\text{s}^*p+P_\text{s}\epsilon^*\}-\mbox{c.c.}), \label{Ptb:D}\\
&\frac{\partial}{\partial t}{p} = -(i\omega_{ba} + \gamma_{\perp}){p}-i\gamma_{\perp}(D_s\epsilon+E_sd), \label{Ptb:P} \\
&\frac{\partial^2 \epsilon}{\partial x^2} - \frac{1}{c^2} \left(\varepsilon_c\frac{\partial^2}{\partial t^2}+\frac{\sigma}{\varepsilon_0} \frac{\partial}{\partial t}\right)\epsilon = \frac{1}{c^2} \frac{\partial^2}{\partial t^2} p \label{Ptb:E}.
\end{align}

Notably, for small $D_\text{p}$, the laser is off, hence $E_\text{s}=0$, $P_\text{s}=0$ and $D_\text{s}=D_\text{p}$. At such a trivial state, equations (\ref{Ptb:D})--(\ref{Ptb:E}) yield linear-cavity wave equations\cite{PhysRevA.82.063824},
\begin{equation}
\frac{\partial^2 \epsilon_m}{\partial x^2} + \frac{\tilde{\omega}_m^2}{c^2}\left(\varepsilon_c+\frac{i\sigma}{\varepsilon_0\tilde{\omega}_m}+\Gamma_\perp(\tilde{\omega}_m)D_\text{p}\right)\epsilon_m = 0 \label{Ln:E},
\end{equation}
where $\Gamma_\perp(\omega)=\frac{\gamma_\perp}{\omega-\omega_{ba}+i\gamma_\perp}$. Equation~\eqref{Ln:E} determines the linear cavity's resonant modes $\epsilon_m$ and the related resonant frequencies $\tilde{\omega}_m$. An EP is approached when two resonant modes merge.

Above the first lasing threshold, a single mode turns on. The solution is a non-trivial fixed point, $E_\text{s}=E_0(x)e^{-i\omega_0t}$, $P_\text{s}=P_0(x)e^{-i\omega_0t}$ and $D_\text{s}=D_0(x)$. Under the stationary-inversion approximation (SIA), Eqs.~(\ref{Ptb:D})--(\ref{Ptb:E}) yield active-cavity modes and determine the second lasing threshold\cite{PhysRevA.82.063824}.


Beyond the SIA, Eqs.~(\ref{Ptb:D})--(\ref{Ptb:E}) determine the comb threshold where single mode lasing becomes unstable to the system becoming a frequency comb.
In this case, the solution is a limit-cycle, 
\begin{align}
\vspace{-2pt}
E_\text{s}(x,t) &= e^{-i\omega_0 t}\sum_{m=-\infty}^{+\infty}E_m(x)e^{-im\omega_\text{d} t},\label{CM:E} \\
P_\text{s}(x,t) &= e^{-i\omega_0 t}\sum_{m=-\infty}^{+\infty}P_m(x)e^{-im\omega_\text{d} t},\label{CM:P} \\
D_\text{s}(x,t) &= \sum_{m=-\infty}^{+\infty}D_m(x)e^{-im\omega_\text{d} t}, \label{CM:D}
\vspace{-2pt}
\end{align} 
where the repetition rate $\omega_\text{d}$, spectral center $\omega_0$ and the Fourier components $\{D_m, E_m, P_m\}$ can all be determined by ``periodic-inversion \textit{ab~initio} laser theory''(PALT) \cite{Gao2024PALT}. 
Eqs.~(\ref{CM:E}--\ref{CM:D}) act as a periodic temporal modulation in Eqs.~(\ref{Ptb:D})--(\ref{Ptb:E}). 
By Floquet theory \cite{annurev-conmatphys-040721-015537}, the solution to Eqs.~(\ref{Ptb:D})--(\ref{Ptb:E}) should include Floquet modes, $f_k(x,t)e^{-i\omega_{\text{F},k}t}$, where $f_k(x,t)$ has a period of $2\pi/\omega_\text{d}$. Due to the difference-frequency generation terms involving $\epsilon^*$ and $p^*$ in Eq.~(\ref{Ptb:D}), each harmonic oscillation $e^{-i\omega_{\text{F},k}t}$ generates a complex-conjugate term $e^{i\omega_{\text{F},k}^*t}$. Hence, we postulate the following form of solutions,
\begin{align}
\vspace{-2pt}
\epsilon(x,t) &= e^{-i\omega_0 t}\sum_k[\epsilon_{ak} e^{-i\omega_{\text{F},k} t} + \epsilon^*_{bk} e^{i\omega_{\text{F},k}^* t}],\label{Fq:E} \\
p(x,t) &= e^{-i\omega_0 t}\sum_k[p_{ak} e^{-i\omega_{\text{F},k} t} + p^*_{bk} e^{i\omega_{\text{F},k}^* t}],\label{Fq:p} \\
d(x,t) &= \sum_k d_{ak} e^{-i\omega_{\text{F},k} t} + d^*_{ak} e^{i\omega_{\text{F},k}^* t}. \label{Fq:d}
\vspace{-2pt}
\end{align} 
Here, $f_{ak,bk}$ with $f=\epsilon, p, d$ are functions of $x$ and $t$, with a time period of $2\pi/\omega_\text{d}$, $f_{ak,bk}=f_{ak,bk}(x,t)=f_{ak,bk}(x,t+2\pi/\omega_\text{d})$. To determine $\omega_\text{F}$ and the space-time profile of $\epsilon_{a,b}(x,t)$, we substitute Eqs.~(\ref{CM:E})-(\ref{CM:D}) and Eqs.~(\ref{Fq:E})-(\ref{Fq:d}) into Eqs.~(\ref{Ptb:D})-(\ref{Ptb:E}). By expanding the periodic functions in Fourier series, we derive the following wave equations,
\begin{align}
\frac{d^2}{d x^2} 
\begin{bmatrix}
\bar{\upepsilon}_{ak} \\
\bar{\upepsilon}_{bk} 
\end{bmatrix}
 + \bm{\Upomega}(\omega_{\text{F},k})
 \begin{bmatrix}
\bar{\upepsilon}_{ak} \\
\bar{\upepsilon}_{bk} 
\end{bmatrix}
 = \mathbf{X}(x,\omega_{\text{F},k})
 \begin{bmatrix}
\bar{\upepsilon}_{ak} \\
\bar{\upepsilon}_{bk} 
\end{bmatrix}, \label{ST}
\end{align} 
in which $\bar{\upepsilon}_{ak,bk}(x)$ is a column vector that includes the Fourier components of $\epsilon_{ak,bk}(x,t)$.
$\bm{\Upomega}(\omega_{\text{F},k})$ and $\mathbf{X}(\omega_{\text{F},k})$ are matrices determined by the limit cycle, $\{E_\text{s},P_\text{s},D_\text{s}\}$. 
$(d^2/dx^2)$ acts element-wise on the column vector. 
$(d^2\bar{\upepsilon}_{ak}/dx^2)$ is associated with outgoing boundaries while $(d^2\bar{\upepsilon}_{bk}/dx^2)$ is associated with incoming boundaries. The derivation of Eq.~(\ref{ST}) and the expressions for $\bm{\Upomega}(\omega_{\text{F},k})$ and $\mathbf{X}(x,\omega_{\text{F},k})$ are provided in the supplementary section I.

Floquet modes can be solved from Eq.~\eqref{ST}. Their Floquet frequencies $\omega_{\text{F},k}$ are determined such that the eigenvectors $\bar{\upepsilon}_{ak,bk}(x)$ are non-trivial. 
For each $\omega_{\text{F},k}$, the expressions of Eq.~\eqref{Fq:E}--\eqref{Fq:d} suggest that $\omega_{\text{F},k'}\equiv-\omega^*_{\text{F},k}$ and $\omega_{\text{F},k''}\equiv\omega_{\text{F},k}+m\omega_\text{d}$ with $m\in \mathbb{Z}$ are degenerate solutions to the same Floquet mode.
Therefore, all the Floquet frequencies can be mapped into the ``Floquet zone'', defined by $\text{Re}(\omega)\in[0, 0.5\omega_d]$ on the complex plane. In the Floquet zone, we define the primary Floquet frequency $\omega_{\text{F}}$ as the $\omega_{\text{F},k}$ with the largest imaginary part, then denote the related mode profile in the expression of Eq.~\eqref{Fq:E} as ${\upepsilon}_{a,b}(x,t)$.
A comb is stable if and only if all perturbations decay over time, which requires $\text{Im}{(\omega_{\text{F},k})}<0$ for all $k$, equivalently $\text{Im}(\omega_\text{F})<0$.

\begin{figure}[h]
\centering
\includegraphics{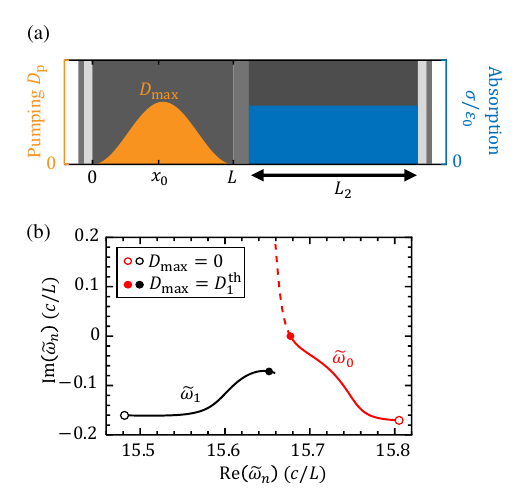}
\vspace{-4pt}
\caption{\textbf{The EP laser with gain-loss coupled cavity}. \textbf{a} A 1D laser cavity with gain on the left and loss on the right. The length of the gain cavity is $L$ and the length of the loss cavity is $L_2 = 1.2L$. The absorption rate inside the loss cavity is $\sigma/\varepsilon_0 = 7.6(c/L)$. The passive refractive index $\sqrt{\varepsilon_c}$ is $3.4$ in the gain cavity and $3.67$ in the passive cavity, typical values of commonly used semiconductor lasing materials. $x_0$ is the location where we plot the field intensity in Fig.~\ref{fig:Dscan} and the phase portraits in Fig.~\ref{fig:LimitCycles}. The values of all passive parameters, including the thickness and the index of each layer, are elaborated in supplementary section II. \textbf{b} the trajectories of two near-EP resonant frequencies solved from Eq.~(\ref{Ln:E}). The first lasing threshold is $D^\text{th}_1 = 1.2$. Dashed lines show the would-be above-threshold trajectories in the absence of gain saturation.
}
\label{fig:System}
\end{figure}

\section{Bistability and Period Doubling Cascade of EP Combs}\label{sec:Example}
\vspace{-4pt}

While the imaginary part of the primary Floquet frequency $\omega_\text{F}$ generally determines the stability of a frequency comb, the real part implies how the comb evolves after it becomes unstable. Specifically, period doubling occurs when (i) $\text{Im}(\omega_\text{F})=0$ and (ii) $\text{Re}(\omega_\text{F})=0.5\omega_\text{d}$. Under condition (i), a random perturbation reduces to a single Floquet mode over time, $\epsilon(x,t) \to e^{-i(\omega_0+\omega_\text{F})t}g(x,t)$, where $g(x,t)=(\epsilon_a+\epsilon_b^*e^{2i\omega_\text{F}t})$. A stronger pumping will lead to the nonlinear effect of wave mixing between the growing Floquet mode and the original comb. If $\text{Re}(\omega_\text{F})/\omega_\text{d}$ is irrational, such wave mixing process will generate an overcomplicated lasing spectrum spreading over all frequencies. However, condition (ii) indicates that $g(x,t+2\pi/\omega_\text{d})=g(x,t)$, hence $\epsilon(x,t)$ consists of Fourier components at $\omega_0+(m+0.5)\omega_\text{d}$, right in the middle of the existing comb lines.  Therefore, wave mixing only generates frequencies of $\omega_0+0.5m\omega_d$, forming another frequency comb with half the repetition rate as before.

We demonstrate such period doubling mechanism in a one-dimensional EP laser cavity
shown in Fig.~\ref{fig:System}a. We adopt a smooth pumping profile $D_\text{p}(x)=0.5D_\text{max}[1-\text{cos}(2\pi x/L)]$ to improve the accuracy of the finite-difference time-domain (FDTD) simulations. Figure~\ref{fig:System}b shows the trajectories of two eigen frequencies solved from Eq.~(\ref{Ln:E}); they are tuned near an EP at the first threshold $D_\text{max}=D^\text{th}_1$, where $\tilde{\omega}_0$ reaches the real axis. Without nonlinear gain saturation, $\tilde{\omega}_0$ would move quickly upward as $D_\text{max}$ increases above $D^\text{th}_1$, while $\tilde{\omega}_1$ would almost stay still, due to the counteraction between being pumped and being repelled by $\tilde{\omega}_0$. 

\begin{figure}[t]
\centering
\includegraphics{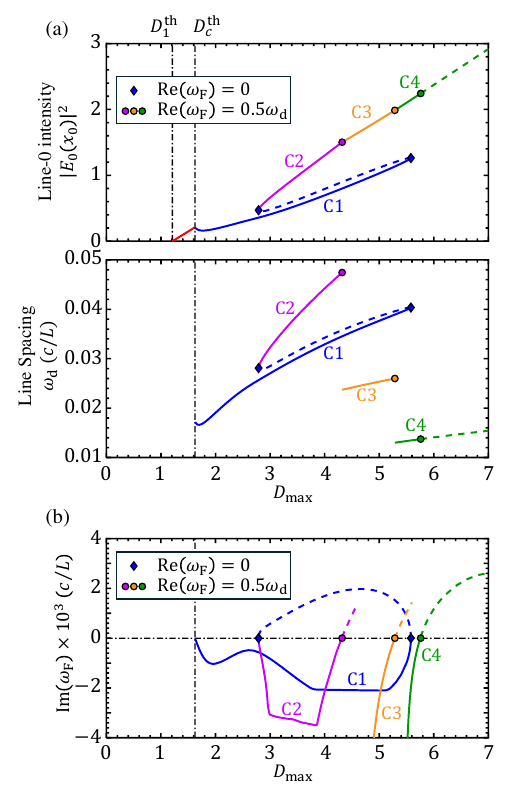}
\vspace{-4pt}
\caption{\textbf{Pump dependence of EP combs}. \textbf{a} PALT calculation of EP combs. (Upper panel) The intensity of central comb line and (lower panel) the repetition rate of EP combs. (Solid lines) Stable solutions and (dashed lines) unstable solutions. (Blue diamonds) the boundaries of bistability. (circles) Period doubling, where $\omega_\text{d}$ drops by half. \textbf{b} The imaginary part of primary Floquet frequencies solved from Eq.~\eqref{ST} for the combs in \textbf{a}.    
}
\label{fig:Dscan}
\end{figure}

For this system, the frequency gap $|\text{Re}(\tilde{\omega}_1-\tilde{\omega}_0)|=0.03(c/L)$ at the first threshold is an estimation to the repetition rate of the EP comb near the comb threshold. It is approximately 30 times smaller than the gain cavity's FSR, $\omega_\text{FSR}=\pi c/\sqrt{\varepsilon_c}\simeq0.9(c/L)$. Without tuning the system closer to the EP, we now show how the repetition can be significantly reduced through period-doubling cascade.
Figure \ref{fig:Dscan}\textbf{a} shows the PALT calculation of the laser states depending on the pumping strength. 
The upper panel shows a continuous transition from the single lasing mode ($D^\text{th}_1<D_\text{max}\leq D^\text{th}_c$) to the major comb line $E_0(x_0)$ across the comb threshold $D^\text{th}_c$. 
The lower panel shows the corresponding repetition rate of the EP combs above $D^\text{th}_c$. Different comb solutions are labeled as ``C''-branches with different colors. 
Fig.~\ref{fig:Dscan}\textbf{b} shows the solutions of primary Floquet frequencies $\omega_\text{F}$ for the EP combs in Fig.~\ref{fig:Dscan}\textbf{a}.  Stable combs are associated with $\text{Im}(\omega_\text{F})<0$. After an $\omega_\text{F}$ crosses the real axis, the corresponding comb transits from one branch to another. 

\begin{figure*}[t]
\centering
\includegraphics[width=1.0\textwidth]{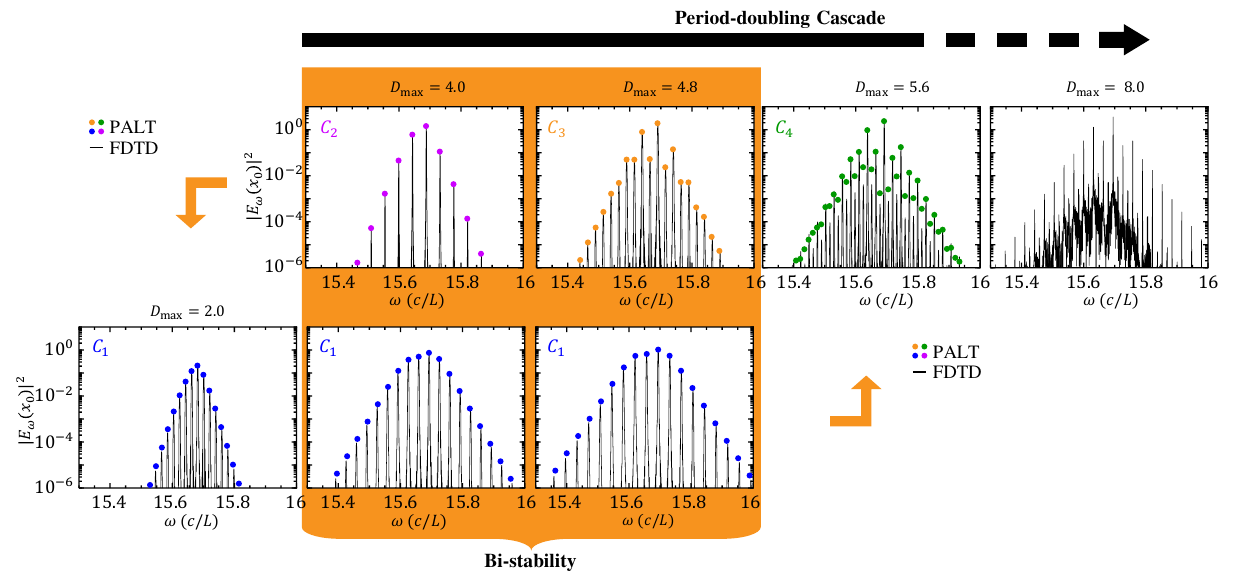}
\vspace{-10pt}
\caption{\textbf{PALT and FDTD simulation of EP-comb spectra at different pumping strengths}. Upper row shows the period-doubling cascade. The bi-stable region is highlighted in orange. Orange arrows show two different paths of how an EP comb evolves. When the pumping $D_\text{max}$ increases regularly from 2.0 to 8.0, the simulation converges to the spectra on the lower row for $D_\text{max}=4.0$ and $4.8$, then jumps to $C_4$ spectrum on the upper row. When the pumping $D_\text{max}$ reduces regularly from 8.0 to 2.0, the simulation converges to the upper row for $D_\text{max}=4.0$ and $4.8$, then jumps down to $C_1$. 
}
\label{fig:Spectra}
\end{figure*}


The filled circles in Fig.~\ref{fig:Dscan}\textbf{a} mark the thresholds of period doubling, where $(\omega_\text{F})=0.5\omega_\text{d}$. 
Above each of these thresholds, the repetition rate reduces by half, shown in the lower panel of Fig.~\ref{fig:Dscan}\textbf{a}.
Figure \ref{fig:Spectra} shows the PALT calculation and FDTD simulation of the lasing spectra at several different pumping strengths selected from Fig.~\ref{fig:Dscan}. The extra comb lines induced by the Floquet mode can be identified on the comb spectrum at $D_\text{max}=4.0$, compared to the spectrum at $D_\text{max}=4.8$. The two spectra also show similar comb bandwidth, implying that the emergence of new comb lines does not compromise the intensity of the existing lines. The period doubling cascade continues after $C_4$ turns unstable (dashed green line with $D_\text{max}>5.76$). Above such a high pumping strength, the theory of PALT and the stability analysis are still valid. However, solving PALT and Eq.~\eqref{ST} becomes excessively time-consuming due to the large amount of Fourier components that must be included to guarantee the accuracy. Despite the difficulty of theoretical analysis, we run FDTD simulations at $D_\text{max}=7$. The simulation result in Fig.\ref{fig:Spectra} shows a continuous spectrum, indicating an infinite number of period doubling procedures occurring within the pumping range of $5.76<D_\text{max}<7$.

In addition to period doubling cascade, our stability analysis also predicts bistable EP combs within the pumping range of $2.8<D_\text{max}<5.6$. In this region, both $C_1$ and the $C_2$--$C_3$--$C_4$ chain are stable. The two stable branches are accessed by different initial conditions. The accessibility is demonstrated by FDTD simulations in Fig.\ref{fig:Spectra}. First, we set the pumping strength to be $D_\text{max}=2.0>D_\text{c}^\text{th}$ and initialize the simulation with a random pulse of the electrical field inside the gain cavity. After the lasing state converges onto the limit cycle, we increase the pumping strength by a small step, then continue the simulation. We iterate such simulation process until $D_\text{max}=8.0$. The simulated comb stays on $C_1$ branch (lower row of the spectra in Fig.\ref{fig:Spectra}), then suddenly jumps up onto $C_4$ (indicated by the right arrow) as $D_\text{max}$ crosses the right end of $C_1$ at 5.6. Second, we start the simulation at $D_\text{max}=8.0$, then regularly reduce the pumping strength. The simulated comb changes backwardly along $C_4 \rightarrow C_3 \rightarrow C_2$, then suddenly jumps down onto $C_1$ as $D_\text{max}$ crosses the left end of $C_2$ at 2.8 (indicated by the left arrow). Thus, the simulation results demonstrate both PALT and the stability analysis on EP combs. 

\begin{figure}[b]
\centering
\includegraphics{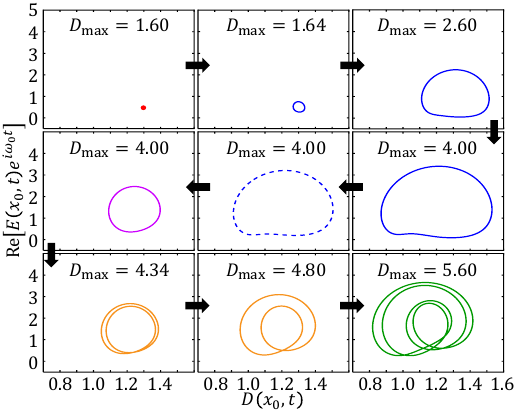}
\vspace{-10pt}
\caption{\textbf{The projection of phase portraits on the plane of population inversion versus electrical field at $x=x_0$}. The horizontal axis is the population inversion $D_s(x_0,t)$ and vertical axis is the envelope of the electrical field $E_s$. Line styles and colors are consistent with Fig.~\ref{fig:Dscan}. 
}
\label{fig:LimitCycles}
\end{figure}

Finally, we summarize the period doubling and bistability phenomena using the phase trajectories of EP combs. The laser's phase space is defined as a manifold with dimensions of $D$ as well as the real and imaginary parts of $E$ and $P$ from Maxwell-Bloch equations Eqs.~\eqref{MB:D}--\eqref{MB:E}. A frequency comb is then recognized as a limit cycle in the phase space\cite{WIECZOREK2005471}. 
For the solutions in Fig.~\ref{fig:Dscan}\textbf{a}, Fig.~\ref{fig:LimitCycles} plots the projections of their phase portraits on the $D(x_0,t)$--Re[$E(x_0,t)$] plane. The red dot in the upper left plot is the single-mode lasing state, known as a fixed point. The first row shows how the fixed point opens and continuously extends into a stable limit cycle. The second row shows the limit cycle shifting from an attractor (right) to a repeller (middle), then reverting to an attractor (left) again. This gives rise to bi-stability. The third row shows period-doubling cascade, where the limit cycle with a basic period doubles its orbit twice. 

\section{Discussion}\label{sec:Example}
In this work, we develop a limit-cycle stability analysis based on Floquet theory. The analysis predicts the novel phenomena of bi-stability and period-doubling cascade of frequency combs in EP lasers. Period doubling cascade occurs when the Floquet frequency crosses real axis at half of the comb's repetition rate. It reduces the repetition rate without shrinking the comb bandwidth, hence allowing for the design of extremely compact OFC generators. The theoretical results are confirmed by FDTD simulations. The four-wave mixing phenomenon observed in a laser diode coupled to a high-Q resonator\cite{Sokol:25} may represent an experimental demonstration of such period-doubling cascade in EP combs. This system possesses all the essential ingredients of an EP: two modes with similar frequencies (one from the laser diode and the other from the high-Q micro-ring cavity), field coupling via backscattering, and gain-loss compensation (gain from the lasing material and loss from the passive cavity).

Our perturbation analysis can be extended to solve scattering problems of periodically-driven nonlinear systems. By including a term of incident wave in Eq.~\eqref{Fq:E}, one can re-derive Eq.~\eqref{ST} with an extra inhomogeneous source term. The Flouqet frequency will then be recognized as $\omega_{\text{in}}-\omega_0$, where $\omega_{\text{in}}$ is the frequency of the incident wave. Future work can study the scattering spectrum of EP lasers by solving such inhomogeneous perturbation equation.

\section{Acknowledgements}
We thank Qinghui Yan for helpful discussions on the phase portraits of limit cycles.
W.W.C.\ acknowledges support from the Laboratory Directed Research and Development program at Sandia National Laboratories.
A.C.\ acknowledges support from the U.S.\ Department of Energy, Office of Basic Energy Sciences, Division of Materials Sciences and Engineering.
This work was performed in part at the Center for Integrated Nanotechnologies, an Office of Science User Facility operated for the U.S. Department of Energy (DOE) Office of Science.
Sandia National Laboratories is a multimission laboratory managed and operated by National Technology \& Engineering Solutions of Sandia, LLC, a wholly owned subsidiary of Honeywell International, Inc., for the U.S. DOE's National Nuclear Security Administration under Contract No. DE-NA-0003525. 
The views expressed in the article do not necessarily represent the views of the U.S. DOE or the United States Government.




\def\bibsection{\section*{\refname}} 
\bibliographystyle{naturemag}
\bibliography{sn-bibliography}

\end{document}


\title{Supplementary Information for Bi-stability and Period-doubling Cascade of Frequency Combs in Exceptional-point lasers}

\maketitle

\onecolumngrid


\thispagestyle{plain} 
\section{Derivation of stability equations}
Here, we derive the equation governing comb stability, as presented in Eq. (14) of the main text, by solving the perturbation equations, Eqs. (4)--(6). In the process, we provide explicit expressions for the entries of the coefficient matrices, $\mathbf{X}(\mathbf{r},\omega_{\text{F},k})$ and $\mathbf{\Upomega}(\omega_{\text{F},k})$. To ensure the generality of the method, the derivation is conducted in three dimensions (3D), after which the result is simplified to obtain Eq. (14) for one dimensional systems. 

The 3D Maxwell--Bloch(MB) equations for electrical field $\mathbf{E}(\mathbf{r},t)$, polarization $\mathbf{P}(\mathbf{r},t)$ and population inversion $D(\mathbf{r},t)$ are
\begin{CAlign}
&\frac{\partial}{\partial t}D = - \gamma_{\parallel}(D-D_\text{p})-\frac{i\gamma_{\parallel}}{2}(\mathbf{E}^*\cdot\mathbf{P}-\mathbf{E}\cdot\mathbf{P}^*), \label{MB:D}\\
&\frac{\partial}{\partial t}\mathbf{P} = -(i\omega_{ba} + \gamma_{\perp})\mathbf{P}-i\gamma_{\perp}D(\mathbf{E}\cdot \hat{\uptheta})\hat{\uptheta}^*, \label{MB:P} \\
-&\nabla\times\nabla\times \mathbf{E} - \frac{1}{c^2} \left(\varepsilon_c\frac{\partial^2}{\partial t^2}+\frac{\sigma}{\varepsilon_0} \frac{\partial}{\partial t}\right)\mathbf{E} = \frac{1}{c^2} \frac{\partial^2}{\partial t^2} \mathbf{P} \label{MB:E}.
\end{CAlign}
$\hat{\uptheta}$ is the unit vector of the atomic dipole momentum, with $\hat{\uptheta}\cdot\hat{\uptheta}^*=1$. The rest of the notation is consistent with Eq. (1)--(3) in the main text. Similar to Eqs. (4)--(6), the 3D perturbation equations can be derived as:
\begin{CAlign}
&\frac{\partial}{\partial t}d = - \gamma_{\parallel}d-\frac{i\gamma_{\parallel}}{2}(\{\mathbf{E}_\text{s}^*\cdot\mathbf{p}+\mathbf{P}_\text{s}\cdot\bm{\epsilon}^*\}-\mbox{c.c.}), \label{Ptb:D}\\
&\frac{\partial}{\partial t}\mathbf{p} = -(i\omega_{ba} + \gamma_{\perp})\mathbf{p}-i\gamma_{\perp}[(D_\text{s}\bm{\epsilon}+\mathbf{E}_\text{s}d)\cdot\hat{\uptheta}]\hat{\uptheta}^*, \label{Ptb:P} \\
-&\nabla\times\nabla\times \bm{\epsilon} - \frac{1}{c^2} \left(\varepsilon_c\frac{\partial^2}{\partial t^2}+\frac{\sigma}{\varepsilon_0} \frac{\partial}{\partial t}\right)\bm{\epsilon} = \frac{1}{c^2} \frac{\partial^2}{\partial t^2} \mathbf{p} \label{Ptb:E},
\end{CAlign}
where $\mathbf{E}_\text{s}$, $\mathbf{P}_\text{s}$ and $D_\text{s}$ solves the Maxwell--Bloch(MB) equations (1)--(3) in the main text. We consider a known limit-cycle solution, 
\begin{CAlign}
\vspace{-2pt}
\mathbf{E}_\text{s}(\mathbf{r},t) &= e^{-i\omega_0 t}\mathbf{E}_\text{env}(\mathbf{r},t) = e^{-i\omega_0 t}\sum_{m=-\infty}^{+\infty}\mathbf{E}_m(\mathbf{r})e^{-im\omega_\text{d} t},\label{CM:E} \\
\mathbf{P}_\text{s}(\mathbf{r},t) &= e^{-i\omega_0 t}P_\text{env}(\mathbf{r},t)\hat{\uptheta}^* = e^{-i\omega_0 t}\sum_{m=-\infty}^{+\infty}P_m(\mathbf{r})e^{-im\omega_\text{d} t}\hat{\uptheta}^*,\label{CM:P} \\
D_\text{s}(\mathbf{r},t) &= \sum_{m=-\infty}^{+\infty}D_m(\mathbf{r})e^{-im\omega_\text{d} t}. \label{CM:D}
\vspace{-2pt}
\end{CAlign} 
where the Fourier components, $\{ \mathbf{E}_m, P_m, D_m \}$, repetition rates, $\omega_\text{d}$ and spectral center $\omega_0$ have been determined by ``periodic-inversion \textit{ab~initio} laser theory''(PALT)\cite{Gao2024PALT}. In this situation, Eq.~\eqref{Ptb:D}--\eqref{Ptb:E} describes a Floquet system under periodic modulation. Therefore, the solutions of $[d,p,\epsilon]$ should be enveloped by Floquet states. Noting the complex--conjugate operation (c.c.) in Eq.~\eqref{Ptb:D}, we consider the trial solution as
\begin{CAlign}
\vspace{-2pt}
\bm{\epsilon}(\mathbf{r},t) &= e^{-i\omega_0 t}\sum_k[\bm{\epsilon}_{ak}(\mathbf{r},t) e^{-i\omega_{\text{F},k} t} + \bm{\epsilon}_{bk}^*(\mathbf{r},t) e^{i\omega_{\text{F},k}^* t}],\label{Fq:E} \\
\mathbf{p}(\mathbf{r},t) &= e^{-i\omega_0 t}\sum_k[p_{ak}(\mathbf{r},t) e^{-i\omega_{\text{F},k} t} + p_{bk}^*(\mathbf{r},t) e^{i\omega_{\text{F},k}^* t}]\hat{\uptheta}^*,\label{Fq:p} \\
d(\mathbf{r},t) &= \sum_k[d_{ak}(\mathbf{r},t) e^{-i\omega_{\text{F},k} t} + d^*_{ak}(\mathbf{r},t) e^{i\omega_{\text{F},k}^* t}]. \label{Fq:d}
\vspace{-2pt}
\end{CAlign} 
\vspace{-2pt}
$\bm{\epsilon}_{a,b}$, $p_{a,b}$ and $d_{ak}$ have the same temporal period of $2\pi/\omega_\text{d}$. The form of Eq.\eqref{Fq:d} ensures that $d$ is real, while $\bm{\epsilon}_{bk}^*$ or $p_{bk}^*$ cannot be assumed as the complex conjugate of $\bm{\epsilon}_{ak}$ or $p_{ak}$, meaning $\bm{\epsilon}_{bk} \ne \bm{\epsilon}_{ak}$ and $p_{bk} \ne p_{ak}$ in general. Now, the target is to determine these periodic envelopes as well as their complex Floquet frequencies $\omega_{\text{F},k}$. First, we substitute Eq.\eqref{CM:E}--\eqref{CM:P} and Eq.\eqref{Fq:E}--\eqref{Fq:d} into Eq.\eqref{Ptb:D} then extract the terms of $e^{-i\omega_{\text{F},k}t}$ for each $k$. To simplify the notation, we now omit the index $k$, as the following derivation applies to all Floquet modes, including the one with the primary Floquet frequency $\omega_\text{F}$.
\begin{CEquation}
    (\frac{\partial}{\partial t}-i\omega_\text{F}+\gamma_\parallel)d_a = -\frac{i\gamma_\parallel}{2}(E_\text{env}^*p_a+P_\text{env}\epsilon_b-E_\text{env}p_b-P_\text{env}^*\epsilon_a), \label{da}
\end{CEquation}
where $E_\text{env}=\mathbf{E}_\text{env}\cdot\hat{\uptheta}$, $\epsilon_b=\bm{\epsilon}_b\cdot\hat{\uptheta}^*$ and $\epsilon_a=(\bm{\epsilon}_a\cdot\hat{\uptheta})$. The $e^{i\omega_\text{F}t}$ component is simply the complex conjugate of Eq.~\eqref{da}. Then, we substitute Eq.\eqref{CM:E}, \eqref{CM:D} and \eqref{Fq:E}--\eqref{Fq:d} into Eq.\eqref{Ptb:P}, then separate the term of $e^{-i(\omega_0-\omega_\text{F}^*)t}$ from that of $e^{-i(\omega_0+\omega_\text{F})t}$. The term of $e^{-i(\omega_0+\omega_\text{F})t}$ yields
\begin{CEquation}
      [\frac{\partial}{\partial t}-i(\omega_0+\omega_\text{F}-\omega_{ba})+\gamma_\perp]p_a=-i\gamma_\perp(D_\text{s}\epsilon_a+E_\text{env}d_a).\label{pa}  
\end{CEquation}
The complex conjugate of $e^{-i(\omega_0-\omega_\text{F}^*)t}$ term yields
\begin{CEquation}
        [\frac{\partial}{\partial t}+i(\omega_0-\omega_\text{F}-\omega_{ba})+\gamma_\perp]p_b=i\gamma_\perp(D_\text{s}\epsilon_b+E_\text{env}^*d_a), \label{pb}
\end{CEquation}
where we have applied the fact that population inversion is real, $D_\text{s}^*=D_\text{s}$. Similarly, we substitute Eq.~\eqref{Fq:E}--\eqref{Fq:p} into the wave equation Eq.~\eqref{Ptb:E}, which yields
\begin{CAlign}
    -&\nabla\times\nabla\times \bm{\epsilon}_a - \frac{1}{c^2} \left[\varepsilon_c(\frac{\partial}{\partial t}-i\omega_0-i\omega_\text{F})^2+\frac{\sigma}{\varepsilon_0} (\frac{\partial}{\partial t}-i\omega_0-i\omega_\text{F})\right]\bm{\epsilon}_a = \frac{1}{c^2} (\frac{\partial}{\partial t}-i\omega_0-i\omega_\text{F})^2 p_a \hat{\uptheta}^*, \label{ea} \\
    -&\nabla\times\nabla\times \bm{\epsilon}_b^* - \frac{1}{c^2} \left[\varepsilon_c(\frac{\partial}{\partial t}-i\omega_0+i\omega_\text{F}^*)^2+\frac{\sigma}{\varepsilon_0} (\frac{\partial}{\partial t}-i\omega_0+i\omega_\text{F}^*)\right]\bm{\epsilon}_b^* = \frac{1}{c^2} (\frac{\partial}{\partial t}-i\omega_0+i\omega_\text{F}^*)^2 p_b^* \hat{\uptheta}^*. \label{eb*}
\end{CAlign}
Before proceeding with Eq.~\eqref{da}--\eqref{eb*}, we introduce the following notation related to Fourier series. For any periodic function $f(t)$, let $f_m$ be it's $m$-th Fourier component. We define $\bar{f}$ as a column vector containing all $f_m$, thus $[\bar{f}]_m = f_m$ for $m\in \mathbb{Z}$. Additionally, we define $\bar{\bar{f}}$ as a matrix, where the $[m,n]$ entry is given by $[\bar{\bar{f}}]_{mn} = f_{m-n}$ for $m,n\in \mathbb{Z}$. In such notation, the multiplication between two functions $f(t)$ and $g(t)$ with the same period can be interpreted as the convolution of their Fourier coefficients, i.e., $\overline{fg} = \bar{\bar{f}}\bar{g} = \bar{\bar{g}}\bar{f}$ and $\overline{f^*g}=\bar{\bar{f}}^\dagger\bar{g}$, where $\dagger$ is the operation of complex-conjugate transpose.

Using the above notation, we expand Eq.~\eqref{da}--\eqref{pb} in Fourier series,
\begin{CAlign}
    \bar{d}_a &= 0.5\mathbf{\Gamma}_\parallel(\bar{\bar{E}}_s^\dagger\bar{p}_a - \bar{\bar{E}}_s\bar{p}_b + \bar{\bar{P}}_s\bar{\epsilon}_b - \bar{\bar{P}}_s^\dagger\bar{\epsilon}_a), \label{dav} \\
    \bar{p}_a &= \mathbf{\Gamma}_a(\bar{\bar{D}}_s\bar{\epsilon}_a+\bar{\bar{E}}_s\bar{d}_a), \label{pav} \\
    \bar{p}_b &= \mathbf{\Gamma}_b(\bar{\bar{D}}_s\bar{\epsilon}_b+\bar{\bar{E}}_s^\dagger\bar{d}_a), \label{pbv}
\end{CAlign}
where $[\bar{\bar{E}}_\text{s}]_{mn}=\mathbf{E}_{m-n}\cdot\hat{\uptheta}$ and $[\bar{\bar{P}}_\text{s}]_{mn}=P_{m-n}$. $\mathbf{\Gamma}_\parallel$, $\mathbf{\Gamma}_a$ and $\mathbf{\Gamma}_b$ are diagonal matrices,
\begin{CAlign} 
[\mathbf{\Gamma}_\parallel]_{mn}&=\gamma_\parallel[m\omega_\text{d}+\omega_\text{F}+i\gamma_\parallel]^{-1}\delta_{mn}, \\
[\mathbf{\Gamma}_a]_{mn}&=\gamma_\perp[\omega_0+m\omega_\text{d}+\omega_\text{F}-\omega_{ba}+i\gamma_\perp]^{-1}\delta_{mn}, \\
[\mathbf{\Gamma}_b]_{mn}&=\gamma_\perp[\omega_0-m\omega_\text{d}-\omega_\text{F}-\omega_{ba}-i\gamma_\perp]^{-1}\delta_{mn},
\end{CAlign}
and $\delta_{mn}$ is Kronecker delta function,  $\delta_{mn} = 1$ for $m=n$, while $\delta_{mn} = 0$ for $m\ne n$. We substitute Eq.~\eqref{pav}--\eqref{pbv} into Eq.~\eqref{dav} to cancel $\bar{p}_a$ and $\bar{p}_b$,
\begin{CEquation}
    \bar{d}_a = 0.5\mathbf{\Gamma}_\parallel[
    (\bar{\bar{E}}_s^\dagger\mathbf{\Gamma}_a \bar{\bar{D}}_s - \bar{\bar{P}}_s^\dagger)\bar{\epsilon}_a - 
    (\bar{\bar{E}}_s\mathbf{\Gamma}_b \bar{\bar{D}}_s - \bar{\bar{P}}_s)\bar{\epsilon}_b +
    (\bar{\bar{E}}_s^\dagger\mathbf{\Gamma}_a\bar{\bar{E}}_s-\bar{\bar{E}}_s\mathbf{\Gamma}_b\bar{\bar{E}}_s^\dagger)\bar{d}_a
    ]. \label{dav2}
\end{CEquation}
Eq.~\eqref{dav2} yields $\bar{d}_a$ as a function of $\bar{\epsilon}_a$ and $\bar{\epsilon}_b$,
\begin{CEquation}
    \bar{d}_a = \bm{\chi}_a \bar{\epsilon}_a - \bm{\chi}_b \bar{\epsilon}_b, \label{dav3}
\end{CEquation}
where 
\begin{CAlign}
    \bm{\chi}_a(\mathbf{r},\omega_\mathbf{F}) &= 0.5[\mathbf{I}-0.5\mathbf{\Gamma}_\parallel(\bar{\bar{E}}_s^\dagger\mathbf{\Gamma}_a\bar{\bar{E}}_s-\bar{\bar{E}}_s\mathbf{\Gamma}_b\bar{\bar{E}}_s^\dagger)]^{-1}\mathbf{\Gamma}_\parallel(\bar{\bar{E}}_s^\dagger\mathbf{\Gamma}_a \bar{\bar{D}}_s - \bar{\bar{P}}_s^\dagger), \\
    \bm{\chi}_b(\mathbf{r},\omega_\mathbf{F}) &= 0.5[\mathbf{I}-0.5\mathbf{\Gamma}_\parallel(\bar{\bar{E}}_s^\dagger\mathbf{\Gamma}_a\bar{\bar{E}}_s-\bar{\bar{E}}_s\mathbf{\Gamma}_b\bar{\bar{E}}_s^\dagger)]^{-1}\mathbf{\Gamma}_\parallel(\bar{\bar{E}}_s\mathbf{\Gamma}_b \bar{\bar{D}}_s - \bar{\bar{P}}_s).
\end{CAlign}

We substitute Eq.~\eqref{dav3} into Eq.~\eqref{pav} and Eq.~\eqref{pbv} to derive $\bar{p}_{a,b}$ as functions of $\bar{\epsilon}_{a,b}$,
\begin{CAlign}
\bar{p}_a &= \mathbf{\Gamma}_a[(\bar{\bar{D}}_s+\bar{\bar{E}}_s\bm{\chi}_a)\bar{\epsilon}_a-\bar{\bar{E}}_s\bm{\chi}_b\bar{\epsilon}_b], \label{pav2} \\
\bar{p}_b &= \mathbf{\Gamma}_b[\bar{\bar{E}}_s^\dagger\bm{\chi}_a\bar{\epsilon}_a+(\bar{\bar{D}}_s-\bar{\bar{E}}_s^\dagger\bm{\chi}_b)\bar{\epsilon}_b]. \label{pbv2}
\end{CAlign}
Eq.~\eqref{pav2}--\eqref{pbv2} imply that the wave equations Eq.~\eqref{ea} and Eq.~\eqref{eb*} are coupled through the source terms, $p_a(\bm{\epsilon}_a,\bm{\epsilon}_b)$ and $p_b(\bm{\epsilon}_a,\bm{\epsilon}_b)$ on the right-hand side. We now expand Eq.~\eqref{ea} and Eq.~\eqref{eb*} in Fourier series,
\begin{CAlign}
    -&\nabla\times\nabla\times \bar{\bm{\epsilon}}_a + \frac{1}{c^2}(\varepsilon_c\bm{\omega}_a^2+i\frac{\sigma}{\varepsilon_0}\bm{\omega}_a)\bar{\bm{\epsilon}}_a = -\frac{1}{c^2} \bm{\omega}_a^2 \bar{p}_a \hat{\uptheta}^*, \label{eav} \\
    -&\nabla\times\nabla\times (\bar{\bm{\epsilon}}_b)^* + \frac{1}{c^2}(\varepsilon_c{\bm{\omega}_b^\dagger}^2+i\frac{\sigma}{\varepsilon_0}{\bm{\omega}_b^\dagger})(\bar{\bm{\epsilon}}_b)^* = -\frac{1}{c^2} {\bm{\omega}_b^\dagger}^2 (\bar{p}_b)^* \hat{\uptheta}^*, \label{eb*v}
\end{CAlign}
where $\bm{\omega}_{a,b}$ are diagonal matrices: $[\bm{\omega}_{a}]_{mn}=(\omega_0+m\omega_\text{d}+\omega_\text{F})\delta_{mn}$ and $[\bm{\omega}_{b}]_{mn}=(\omega_0-m\omega_\text{d}-\omega_\text{F})\delta_{mn}$. The spatial derivative operator ($-\nabla\times\nabla\times$) acts element-wisely on each Fourier component of $\bm{\epsilon}_{a,b}$. To combine Eq.~\eqref{pav2}--\eqref{eb*v} into a group of linear equations, we must take the complex conjugate of Eq.~\eqref{eb*v}, which also flips the boundary conditions. In doing so, we derive the stability equations for the limit cycle,
\begin{CAlign}
 -&\nabla\times\nabla\times 
\begin{bmatrix}
\bar{\bm{\upepsilon}}_a \\
\bar{\bm{\upepsilon}}_b 
\end{bmatrix}
 + \bm{\Upomega}(\omega_\text{F})
 \begin{bmatrix}
\bar{\bm{\upepsilon}}_a \\
\bar{\bm{\upepsilon}}_b 
\end{bmatrix}
 = \mathbf{X}(\mathbf{r},\omega_\text{F})
 \begin{bmatrix}
\hat{\uptheta}^*(\hat{\uptheta}\cdot\bar{\bm{\upepsilon}}_a) \\
\hat{\uptheta}(\hat{\uptheta}^*\cdot\bar{\bm{\upepsilon}}_b) 
\end{bmatrix}, \label{ST}
\end{CAlign} 
where 
\begin{CAlign}
 \bm{\Upomega} &= \frac{1}{c^2}\varepsilon_c \left(
\begin{bmatrix}
\bm{\omega}_a & \mathbf{0}\\
\mathbf{0} & \bm{\omega}_b 
\end{bmatrix}^2
+i\frac{\sigma}{\varepsilon_0}
\begin{bmatrix}
\bm{\omega}_a & \mathbf{0}\\
\mathbf{0} & \bm{\omega}_b 
\end{bmatrix} \right),\\
\mathbf{X} &= -\frac{1}{c^2}
\begin{bmatrix}
\mathbf{\Gamma}_a\bm{\omega}_a^2  & \mathbf{0}\\
\mathbf{0} & \mathbf{\Gamma}_b\bm{\omega}_b^2
\end{bmatrix}
\begin{bmatrix}
\bar{\bar{D}}_s+\bar{\bar{E}}_s\bm{\chi}_a & -\bar{\bar{E}}_s\bm{\chi}_b\\
\bar{\bar{E}}_s^\dagger\bm{\chi}_a & \bar{\bar{D}}_s-\bar{\bar{E}}_s^\dagger\bm{\chi}_b 
\end{bmatrix}.
\end{CAlign}
In one-dimensional system where $\partial_y = \partial_z = 0$ and $\hat{\uptheta}=\uptheta\hat{z}$, the perturbation of the electrical field can be simplified as $\bar{\bm{\epsilon}}_{a}=\bar{{\epsilon}}_{a}\uptheta^*\hat{z}$ and $\bar{\bm{\epsilon}}_{b}=\bar{{\epsilon}}_{b}\uptheta\hat{z}$. Thus, Eq.~\eqref{ST} reduces to Eq. (14) in the main text,
\begin{CAlign}
\frac{d^2}{d x^2} 
\begin{bmatrix}
\bar{\upepsilon}_{ak} \\
\bar{\upepsilon}_{bk} 
\end{bmatrix}
 + \bm{\Upomega}(\omega_{\text{F},k})
 \begin{bmatrix}
\bar{\upepsilon}_{ak} \\
\bar{\upepsilon}_{bk} 
\end{bmatrix}
 = \mathbf{X}(x,\omega_{\text{F},k})
 \begin{bmatrix}
\bar{\upepsilon}_{ak} \\
\bar{\upepsilon}_{bk} 
\end{bmatrix}. \label{ST:1D}
\end{CAlign} 
where we have added back the index $k$.

\section{Passive parameters of the EP laser}
\begin{figure}[h]
\centering
\includegraphics[width=0.75\textwidth]{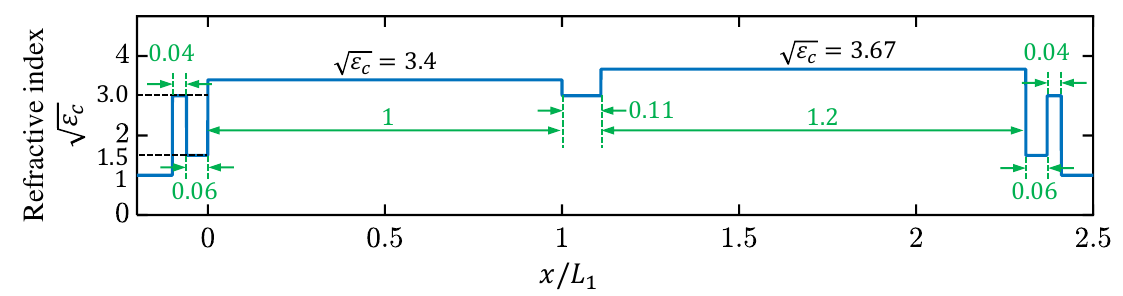}
\vspace{-4pt}
\caption{\textbf{The spatial profile of the refractive index in the EP laser}. 
\label{Fig:index}
}
\end{figure}

Supplementary Fig.\ref{Fig:index} shows the passive refractive index of the EP laser cavity in the main text. The spectra in Fig.\textbf{3} and the phase portraits in Fig.\textbf{4} of the main text are recorded inside the gain cavity at $x_0=0.47L$. 



\clearpage

\bibliographystyle{naturemag}
\bibliography{sn-bibliography}